# Photon statistics analysis of h-BN quantum emitters with pulsed and continuous-wave excitation

Hamidreza Akbari[1], Pankaj K. Jha[1,2], Kristina Malinowski[1], Benjamin E.C. Koltenbah[3], and Harry A. Atwater[1]

[1]*Thomas J. Watson Lboratory of Applied Physics, California Institute of Technology, Pasadena, CA., 91125*

[2] *Department of Electrical Engineering and Computer Science, Syracuse University, Syracuse, NY., 13244*

[3]*Boeing Research & Technology, P.O. Box 3707 MC 1M-201, Seattle, WA. 98124*

**Abstract**

We report on the quantum photon statistics of hexagonal boron nitride (h-BN) quantum emitters by analyzing the Mandel Q parameter. We have measured the Mandel Q parameter for h-BN quantum emitters under various temperatures and pump power excitation conditions. Under pulsed excitation we can achieve a Mandel Q of -0.002 and under continuous-wave (CW) excitation this parameter can reach -0.0025. We investigate the effect of cryogenic temperatures on Mandel Q and conclude that the photon statistics vary weakly with temperature. Through calculation of spontaneous emission from an excited two-level emitter model, we demonstrate good agreement between measured and calculated Mandel Q parameter when accounting for the experimental photon collection efficiency. Finally, we illustrate the usefulness of Mandel Q in quantum applications by the example of random number generation and analyze the effect of Mandel Q on the speed of generating random bits via this method.

**Introduction**

Hexagonal boron nitride (h-BN) is a 2D Van der Waals semiconductor with an energy gap of 6 eV [1]. Point defects in h-BN can form color centers that act as single photon sources [2–4]. The underlying atomic structure of the class of h-BN emitters emitting at around 585 nm (employed in this study) is reported to be associated with carbon impurities [5], but the exact atomic structure has not been definitively identified [5]. Quantum light emitters in hBN are highly stable at room temperature [6] and also at higher temperatures up to 800K [7] with a large Debye-Waller factor (>80%) [2]. These emitters exhibit tunable [8–10] and lifetime limited linewidth at cryogenic temperatures [10,11] and reportedly also at room temperatures [12]. Due to the properties of the hexagonal boron nitride host crystal, these emitters are promising candidates for on chip [13] quantum-nano-photonic applications such as quantum sensing and quantum computation [14].

Many quantum light emission applications require single photon emission (preferably an indistinguishable single photon) on demand [15]. To satisfy this requirement one needs to ultimately achieve sub-Poissonian photon statistics. In other words, if we divide the photon generation into time intervals, there needs to be one and only one photon generated per interval. With this vision in mind, Mandel proposed the notion of the Q number [16]. Q is defined as:

$$Q = \frac{S^2{}_N}{\langle N \rangle} - 1 \qquad (1)$$

Here $S^2{}_N$ is the variance of the photon number per interval and $\langle N \rangle$ is the average number of photons per interval. As shown in Fig. 1 we can divide the stream of photons into several time-bins and measure the statistics of number of photons per bin. For a coherent source of light (laser) the number distribution follows a Poissonian distribution [17], in which the average and variance are equal, hence $Q_{LASER}=0$. For an ideal single photon source there is 1 and only 1 photon per interval so the variance is zero, hence $Q_{Ideal\ SPE}=-1$. This mode of statistics where variance is less than that of a Poisson distribution is called sub-Poissonian. For a thermal source of light, the statistics are super-Poissonian, i.e., $Q > 0$ [17]. In classical electromagnetic theory, $Q > 0$ corresponds to a continuous beam of light which exhibits variable intensity, which is in accord with the bunching phenomena in the quantum picture and a high variance in the statistics.

While an ideal SPE demonstrates $Q = -1$, under experimental conditions with non-unity photon collection efficiency, measurements of SPEs result in $-1 < Q < 0$. The absolute value of Q for each excitation cycle is limited by the emission efficiency of the SPE and experimental photon collection efficiency [18]. Mandel Q has been studied previously, mainly under pulsed excitation [17, 18]. Theoretically the single pulse Mandel Q is shown to be correlated with the value of $g^{(2)}(0)$ [17]. However, the experimental value of $Q_{single\ pulse}$ is limited by the collection efficiency of the experimental setup [18]:

$$Q_{experimental} = Q_{single\ pulse} \times \eta \qquad (2)$$

where $\eta$ is the combined losses in the photon detection process. Both $g^{(2)}(0)$ and Mandel Q can be used to confirm the quantum nature of the emitter. However, as seen in equation (2) Mandel Q depends on the collection efficiency of the optical setup, whereas $g^{(2)}(0)$ does not. The difficulty of measuring the $Q_{single\ pulse}$ due to losses have likely discouraged authors of previous reports to measure this parameter, and thus most only report the $g^{(2)}$ function to indicate the quantum nature of the emission. Furthermore, as we also demonstrate in this study, Mandel Q can be a useful indicator of the utility of emitters in applications that require on demand single photon sources or a high rate of single photon generation. This idea stems from the fact that as Q increases from -1 to 0, the probability of generating a single photon decreases, and instead the probability of not generating a photon or generating multiple photons increases. If an application relies on generating a certain number of single photons, then can be done more quickly with an emitter exhibiting Q closer to -1.

Photon statistics for quantum emitters under CW excitation have been less well studied. We are aware of only one report under such conditions [19], which concluded that the single photon detector dead time plays a major role in determining the photon statistics. Most applications cited for quantum emitters rely on pulsed excitation owing to the degree of temporal control of photon generation and detection. However, an extensive study of CW Mandel Q can help in assessing the use quantum emitters for applications under CW excitation. We show that it is possible to measure the sub-Poissonian statistics under CW excitation with an appropriate method for tagging and binning single photons.

Here, we investigate the Mandel Q parameter using the traditional method of pulsed excitation and also explore the temperature dependence of pulsed Mandel Q, supporting our measurements with simulations. Then, we explore the photon statistics under CW pumping by a laser source, generating a stream of single photons. In this regime, the natural time scale for photon pumping and decay dictates the emission statistics. We show that under CW pumping we can improve upon the inherent efficiency limit of our detection setup for Mandel Q and that pump power can be used as knob to tune Mandel Q. We use this to further demonstrate the usefulness of Mandel Q in random number generation applications that require on-demand single photons.

**Pulsed Mandel Q Measurement**

Our analysis employs single photon emitters in exfoliated flakes of h-BN on a Si/SiO$_2$ substrate. The flakes are scanned in our homebuilt confocal microscope to find single photon emitters: each emitter observed to exhibit a spectrally narrow zero phonon line feature is subject to intensity autocorrelation measurements to measure the g$^{(2)}$ function. For a particular emitter studied here, we observe a zero-phonon line at 588 nm and a $g^{(2)}(0) = 0.07$ that corresponds to a single photon emitter. The emission rate shows two distinct "ON" and "OFF" states switching between 50 to 100 kHz emission rate. This phenomenon is known as blinking [18, 20] and has been proposed to result from the existence of a metastable third state in the energy level configuration of the emitter (as shown in Fig. 2-g) . Pulsed g$^{(2)}$ measurements exhibit antibunching, and the polarized emission pattern is consistent with a dipole radiation pattern matching other studies [21,22]. We calculated the Mandel Q of this emitter by measuring the statistics of photons generated by each pulse in the time span between two consecutive pulses. To remove some of the background noise, we performed time gating [18] to remove photons arriving at times longer than 10ns (~10 times the decay lifetime of the emitter). Based on the photon losses for our microscope (efficiency ~0.005) we expect most of the pulses to show zero counts and some pulses containing 1 or two photons. We don't expect to observe more than 2 photons per pulse in this regime as it is limited by the dead time of the detectors. We chose the pulse-to-pulse duration of 100 ns based on the deadtime of our avalanche photodiode single photon detectors (see methods section).

To compare our results with both coherent and thermal light sources, we performed experiment with CW laser light and thermal light generated by an incandescent light bulb. As can be seen in Table 1, we observe a $Q = (-1.97 \pm 0.42) \times 10^{-3}$ for this h-BN emitter. We can confirm this is Q < 0, within the limits of measurement error for data derived from averaging the results of 5 experiments. For laser light $Q = (-1.25 \pm 4.01) \times 10^{-4}$, which corresponds to a Mandel value of zero within measurement error; the slightly negative value can be explained by the effect of dead time on the measurement of two consecutive photons. Every instance recorded with two photons is a result of measuring one photon at channel 0 and another photon at channel 1. Measurement of two consecutive photons on one channel is not possible because after recording the first instance, the channel will enter a dead time in which it cannot record another photon count. As a result, the number of two photon events that are recorded is smaller than the actual number of two photon events, resulting in a slightly negative Mandel Q. For thermal light we observe $Q = $ $(6.59 \pm 0.77) \times 10^{-3}$ which is positive within measurement error, as expected. Fig. 2-h compares the Mandel Q of the single photon source, laser source, and thermal light source.

**Effect of temperature on Mandel Q**

We performed photon statistics measurements for another emitter prepared via the same method on the SiO$_2$ on Si substrate in a cryogenic microscope (see Methods) to study the effect of temperature on the photon statistics. This emitter also exhibited narrow zero phonon line at 585nm (see Fig. 3-a), where this emission energy has been attributed to carbon impurities in the h-BN crystal [5]. This emitter exhibited $g^{(2)}(0) = 0.15$ which confirms the quantum nature of light emission. The g$^{(2)}$ function (Fig. 3-b) also shows a large bunching shoulder corresponding to a three level system. To measure the true linewidth of the emission we performed photoluminescence excitation spectroscopy experiments in which we spectrally scan through the zero-phonon line with a tunable CW laser (<100 kHz linewidth) and measure the emission rate in the sideband (630nm-650nm). Fig. 3-c shows the result of this measurement, which yields a linewidth of 305 MHz. Even though this emitter does not exhibit a fully lifetime-limited linewidth, it nonetheless very little spectral diffusion broadening at 7 K.

To study the effect of temperature on hBN emitter photon statistics, we measured the Mandel Q for temperature values of 7K, 50K, 100K, 150K, 200K, 250K. For each temperature we excited the emitter with pulsed 532nm laser and performed the same time gating procedure described above to minimize the background noise. The result in Fig. 3-d shows that while the minimum Mandel Q value is achieved at 7K, the correlation with temperature is not strong and the value of Mandel Q fluctuates around zero with a minimum of -0.0002±0.0005. All the values measured are consistent with a value of zero within measurement error. We estimate the optical efficiency to be around 0.0005 in our cryo-microscope, which can explain the lower values of Mandel Q in low temperature measurements as compared with room temperature experiments.

**Simulation**

The pulsed Mandel Q parameter measurements were compared with a simplified model of a two-level emitter system. We utilized an extended version of the Jaynes-Cummings Model (eJCM), in which we defined multiple field modes into which an excited emitter may spontaneously emit. The field modes represent the discrete Fourier transform of a temporal pulse with duration commensurate with the lifetime of the hBN emitter. Utilizing the eJCM in this manner defines a system of pseudo-periodic emission and re-absorption of a single photon not unlike the familiar Rabi oscillations of single mode JCM interaction. However, our multi-modal interaction demonstrates averaged emission of a single spontaneous pulse followed by pseudo-stability over the time frame of one such oscillation, within which time we calculate photon statistics and observe near-perfect single photon emission denoted by a Mandel Q value of nearly -1 for each pulsed spontaneous emission event.

We recognize that restriction of emission modes to just those of a narrow band and single direction propagation of an emitted pulse does not incorporate all the emission modes and environmental interactions that would be more comprehensively descriptive of the hBN system. We are presently developing such a description better-suited to such physical interactions, based on the following eJCM derivations, but extended with wave packet, input-output and master equation formalism such as derived and utilized in [23–25]. However, we note similarity between our present simplified method and the limit of no environmental interaction in Appendix A in [23] for the case of spontaneous emission from a two-level emitter, again with restriction of our results to within the first periodic interaction. We therefore deem our present simplified treatment sufficient for these qualitative and limited quantitative Mandel Q comparisons between measurement and calculation.

Brief details of the calculation are as follows. We envision an emitter as a radiative dipole with spontaneous emission as a temporal pulse. We define discrete field modes of varying frequency about the transition frequency of the emitter by first presuming a nominally Gaussian temporal field pulse envelope shape with parameter $\sigma = 0.5$ ns, total pulse duration of 8 ns, and then discretize the function into $N_f$ points in time as shown in Fig. 4-a, which comprises discrete points from several number of modes $N_f$ up to 11. The origin of this Gaussian pulse shape is from our typical definition of an initial *illuminating* pulse that at least partially *stimulates* emission in our model, however in this present case the initial occupation of these modes is zero as we are first examining strictly *spontaneous* emission, although this will be modified below to introduce photon noise into the calculations, at least in a simplified manner. The field mode definitions, however, remain the same in either case, and we utilize these modes for both emission and any initial photon noise. Through discrete Fourier transform, we then define $N_f$ field modes in frequency. The resulting modes have narrow band frequencies around the central frequency of the two-level transition, which is defined with wavelength 588 nm, matching the experimental measurements. Figure 4-b shows normalized partition of pulse power (blue) into these frequency modes (for $N_f = 11$), again corresponding to the nominal Gaussian distribution in time from Fig. 4-a. The field modes are also assumed linearly polarized in direction parallel to the assumed dipole moment of the emitter.

The two-level standard single-mode interacting JCM Hamiltonian [26] is extended to include $N_f$ multiple field mode interaction as follows:

$$\hat{H} = \frac{\hbar\omega_o}{2}(\hat{\sigma}_{2,2} - \hat{\sigma}_{1,1}) + \sum_{j=1}^{N_f} \hbar\omega_j \, \hat{a}_j^\dagger \hat{a}_j - \hbar\omega_o \sum_{j=1}^{N_f} \gamma_j \left(e^{i(\omega_j - \omega_o)t}\hat{a}_j^\dagger \hat{\sigma}_{1,2} - e^{-i(\omega_j - \omega_o)t}\hat{a}_j \hat{\sigma}_{2,1}\right)$$
(3)

The eJCM Hamiltonian is expressed in the rotating wave approximation, where fast oscillating interaction terms on the order of the optical transition frequency have been dropped, and terms of slower oscillation on the time frame of the pulse duration (frequency difference terms) remain. In Eq. (3), with two-level ground and excited emitter states $|g\rangle$ and $|e\rangle$, respectively, the emitter operators are $\hat{\sigma}_{1,1} = |g\rangle\langle g|$, $\hat{\sigma}_{2,2} = |e\rangle\langle e|$, $\hat{\sigma}_{2,1} = |e\rangle\langle g|$ and $\hat{\sigma}_{1,2} = |g\rangle\langle e|$. $\omega_o$ is the resonance frequency of the emitter transition, $\omega_j$ are the modal frequencies, closely centered around $\omega_o$. $\hat{a}_j^\dagger$ and $\hat{a}_j$ operate on photon number states from $N_f$ orthogonal Hilbert spaces, where $[\hat{a}_i, \hat{a}_j^\dagger] = \delta_{i,j}$. $N_f$ is conveniently defined as an odd number with modes arranged in increasing frequency order and middle value on resonance with $\omega_o$. $\gamma_j$ is the normalized modal coupling between photon Mode $j$ and the emission dipole. This coupling is frequency dependent, although all the interaction coefficients used here are quite close in value due to the narrow band of modes. Its value corresponds to a Rabi oscillation time of about 6 ns as when the emitter interacts with a single resonant mode, although it will be shown that initial emission for the multi-modal case corresponds to a more rapid emission on par with the measured lifetime of the emitter. This set the baseline for the calculations, and all multi-modal results were examined within this first Rabi fundamental oscillation time.

The density matrix $\hat{\rho}(t)$, describing the time evolution of the system of states, is therefore comprised of $N_f$ Hilbert spaces of number states for the photon modes and a Hilbert space for the two-level emitter states described before. The initial density matrix ($\hat{\rho}(t=0) = \hat{\rho}_o$) comprises, for our first set of calculations, zero photons in each mode and fully excited emitter with occupancy of the upper state $|e\rangle$ as $p_E = 1$. Additional calculations were then made through a sweep of emission efficiency values, or upper state occupation, of $p_E = 0 – 1$. We also introduced thermal noise into the initial state through a simple fashion by defining a finite average number of thermal photons $N_o$ partitioned into the $N_f$ modes using thermal statistics, calculating a sweep of $N_o = 0 – 1$. Since this calculation is only comprised of a single photon emission and no more than 1 photon is initially distributed through all the modes, it is sufficient to limit the number of photon states per mode to 2, although 3 was used as a matter of increased accuracy to adequately define initial thermal distributions for $N_o > 0$.

We transform the density matrix in time using a unitary transform:

$$\hat{\rho}(t) = \hat{U}(t)\hat{\rho}_o\hat{U}^\dagger(t) \quad (4)$$

The transform $\hat{U}(t)$ is solution of the following equation of motion [27]:

$$\frac{d\hat{U}(t)}{dt} = -\frac{i}{\hbar}\hat{H}(t)\hat{U}(t) \quad (5)$$

Upon projection of Equation (5) onto the orthogonal set of states for the $N_f$ photon and two-level Hilbert spaces, we generate a system of coupled ordinary differential equations (ODEs), expressed in multi-dimensional matrix representation. The initial conditions of the unitary transform equate it with the identity operator as $\hat{U}(t=0) = \hat{I}$. Representing all the modal photon and emitter projection states in shorthand as $|\alpha\rangle, |\beta\rangle, |\gamma\rangle$, we can express (3) as a system of ODEs as follows, using respective collective matrix indices $\alpha, \beta, \gamma$:

$$\frac{d}{dt}U_{\alpha,\beta}(t) = -\frac{i}{\hbar}\sum_\gamma H_{\alpha,\gamma}(t)U_{\gamma,\beta}(t) \quad (6)$$

From the transformed density matrix $\hat{\rho}(t)$, we calculate the average number of photons in each mode as well as each mode's square number average. From these we calculate both the modal and the full collective pulse statistics and, finally, the Mandel Q parameter, all as function of interaction time $t$. We calculate the following statistical results for each photon mode:

$$N_j(t) = Tr[\hat{\rho}(t)\hat{a}_j^\dagger\hat{a}_j], \quad N_j^2(t) = Tr[\hat{\rho}(t)(\hat{a}_j^\dagger\hat{a}_j)^2], \quad N_{i,j}^2(t) = Tr[\hat{\rho}(t)\hat{a}_i^\dagger\hat{a}_i\hat{a}_j^\dagger\hat{a}_j] \quad (7)$$

The full pulse statistics can then be calculated from Eqs. (7) as:

$$N(t) = \sum_{j=1}^{N_F} N_j(t), \qquad N^2(t) = \sum_{j=1}^{N_F} N_j^2(t) + 2\sum_{i=1}^{N_F}\sum_{j=1}^{i-1} N_{i,j}^2(t) \qquad (8)$$

$$var(t) = N^2(t) - (N(t))^2, \qquad Q(t) = \frac{var(t)}{N(t)} - 1$$

The results of the collective mode statistics from Eq. (8) are shown in Fig. 4-c as function of interaction time for $N_f = 5$ for four cases: Case 1 (blue) $p_E = 1$, $N_o = 0$; Case 2 (red) $p_E = 0.7$, $N_o = 0$; Case 3 (green) $p_E = 1$, $N_o = 0.5$; Case 4 (purple) $p_E = 0.7$, $N_o = 0.5$. The figure shows the average emitted photon number $N(t)$ and corresponding $Q(t)$ parameter through an initial rapid partial emission followed by longer stabilized emission corresponding to the size of the defined pulse length within several $\sigma$ values. For Case 1, Mandel Q reaches near perfect single photon statistics of -1 from about $t = 3 - 5$ ns before the interaction time nears the model's periodicity and undergoes reabsorption of the photon to be followed by another emission. Again, we state that the first pulsed emission can be effectively treated as if in isolation within this first time period, and this is born out in Fig. 4-c. With addition of both initial thermal photon noise and inefficiency in the initial excitation of the emitter, the $Q$ value decreases in absolute value, trending closer to zero. Cases 3 and 4, with finite initial noise, both show initial upward trend in the Q value indicative of thermal statistics prior to the full interaction resulting in fractional photon emission and negative-valued Q, albeit affected by intermixing with the noise.

The resulting photon occupation of the modes, taken at optimal interaction time $t_o = 4$ ns, were used to construct the modal occupations of the spontaneously emitted pulse, shown in red in Fig. 4-b. Compared with the presumed Gaussian distribution shown in blue, the actual calculated distribution is fit by a much narrower and higher-peaked Lorentzian function around the central wavelength. (Note that, for better clarity, the calculations of Fig. 4-b were done with $N_f = 11$, $p_E = 1$ and $N_o = 0$, where its results vs. interaction time were identical to those from Case 1 in Fig. 4-c.)

To better examine the effects of varying both the emitter efficiency and initial thermal noise, the photon statistical results are shown in Figs. 5-a and 5-b, respectively, all extracted at the optimal interaction time $t_o = 4$ ns for $N_f = 5$. In Fig. 5-a, $N_o = 0$, and both the number of emitted photons and Q values are linear with $p_E$, varying from the trivial values of 0 for $p_E = 0$ (no spontaneous emission) to 1 emitted photon and Q = -1 (complete single photon emission). In Fig. 5-b, the statistics are shown with varying number of average thermal photons using quantum efficiencies of 1 and 0.7, the latter of which was previously measured for the hBN emitter [22,28]. As expected, both the average number of photons and variance after interaction increase with the added photon noise, the Q values decrease in magnitude towards zero, and the decrease in emission efficiency reduces the number of photons, increases the variance, decreases the magnitude of the Q value as expected.

Taking into account the aforementioned measurement inefficiency of 0.005, we added an imbalanced beam splitter following the emitter interaction with ratio of energy transmission to reflection of (T:R) = (0.005:0.995) to collectively represent other presumed sources of loss, inefficiencies, etc., in a simplified fashion. Transformation of the photon density matrix through the beam splitter results in photon statistics and subsequent Mandel Q value with best (most negative) value on the order of $-5 \times 10^{-3}$, a value commensurate with the pulsed measurements of about -2 x 10$^{-3}$ as well as the estimated effects of Eq. (*2*). Repeating the plot of statistical results from Fig. 5-b, we show in Fig. 5-c the corresponding results, now following transformation through the high-loss beam splitter. Variance values are now slightly smaller than their corresponding average photon number values, resulting in small-valued, but negative Q values. All trends with varying initial noise and emitter efficiencies remain the same as before the loss transformations.

Despite the simplified model with simplified application of noise, emitter efficiency and simulated measurement loss, the favorable comparison of Mandel Q between calculated and measured values serves to further indicate the pulsed emission from the hBN emitter is indeed behaving as a single photon source, which, accounting for experimental inefficiencies, can be represented even semi-quantitatively by a simple two-level emitter model of pulsed emission.

**Photon statistics with continuous wave (CW) excitation**

We also studied the photon statistics of an h-BN emitter in the CW regime. With CW pumping, we have the freedom to choose the time bin length for calculating Mandel Q as there is no intrinsic time scale (other than the lifetime of the emitter). This concept is depicted in Fig. 6-a. However, as noted previously [19], detector dead time can skew the result towards more negative numbers. To resolve this issue, we have limited measurements to a regime in which the inverse count rate of photons (~1000 ns) is much longer than the dead time (~100 ns). Thus, the average time span between two consecutive photons is much longer than the dead time, minimizing the effect of dead time in our results.

We examined a range of time bin sizes, optimizing the time bin to obtain the most negative value of Mandel Q, as shown in Fig. 6-b. We denote this value as $Q_{min}$ and the corresponding time bin size as $T_{Qmin}$. The time bin size in which the emitter reaches Q=0 is also of note. We call this time scale the onset of coherence, as emitters at this condition exhibit Poissonian photon statistics. Increasing the bin size any further produces super-Poissonian statistics. Further, we use pump power as a knob to change the photon emission rate and study the CW photon statistics of the h-BN quantum emitter. Fig. 6-c shows the power is varied in the linear regime of excitation-emission and not close to saturation. However, we notice a non-monotonic dependence of minimum Mandel Q on pump power (see Fig. 6-d).

To understand the non-monotonic dependence of minimum Q on pump power, we investigated the blinking behavior of the emitter at various pump powers and performed measurements so as to isolate the "ON" times (in which the emitter is in a brighter state) and "OFF" times (in which the emitter is in a darker state). We then calculate Mandel Q for each case individually. Fig. 7 shows that when calculating the Mandel Q solely for times when the emitter is "ON", the Minimum Mandel Q is inversely correlated with the pump power. We want to note at the maximum pump power the emitter became very unstable and was only in the "ON" state briefly. Therefore, the observed non-monotonic dependence of minimum Mandel Q on pump power (at all times), was a result of varied ratios of "ON" to "OFF" times at different powers.

**Quantum Random Number Generation and Mandel Q**

Since we can use pump power to tune the CW Mandel Q, we used this advantage to demonstrate an exemplary use case for Mandel Q measurement in random number generation. Random number generation is a quantum application that depends on on-demand singe photon sources. We note that other studies have already used h-BN emitters for random number generation [29,30]. We propose two protocols to generate random bits from h-BN single photon emitter to establish the usefulness of Mandel Q in the demonstrating randomness of the bits generated. In the first method, we use each individual photon recorded in the two single photon detectors of our Hanbury Brown-Twiss setup. If the photon was detected at channel 0, we generated a bit with value 0 and if the photon was recorded at channel 1, we generated a bit with value 1. In the second method, we chose a time bin size equal to $T_{Qmin}$ and bin the detected stream of photon events based on this value. Then we search for bins with 1 and only 1 photon in the bin. Finally, if the singular photon is recorded in channel 0, we generated a bit with value 0 and if the photon is in bin is recorded in channel 1, we generated a bit with value 1. (See details about random number generation in the Supporting Information).

To compare these protocols, we analyzed the random numbers generated using the NIST randomness test [31]. This test suite performs 16 different tests to characterize the randomness of the bits generated. With each method, we generated 400,000 bits, and based on the result shown in Fig. 8-c, the bits generated with method 1 did not pass 4 of the 16 tests, including 'Run test", "Longest run of ones in a block", "Overlapping template matching test", and "Approximate entropy test". However, the bits generated by method 2 passed all the tests. We note here that in the NIST randomness test suite, emphasis is placed simply on passing the threshold for randomness, not on the significance magnitude.

Furthermore, we calculated the time it took to generate one bit on average for each measured pump power and showed that, as CW Mandel Q decreases, the random bit generation rate increases. This is an intuitive conclusion as we know the most negative Q=-1 means the source is generating one photon per cycle and, as Q continues to increase towards zero, there is higher and higher uncertainty in the photon generation process. As a result, with a more negative value of Q, the process can be performed in fewer cycles. This result demonstrates the use of Mandel Q for quantum applications that require on demand single photons, such as quantum communication and quantum computation.

**Conclusion**

In summary, the Mandel Q parameter is an appropriate and useful measure of the photon statistics of h-BN emitters under both pulsed and CW excitation, at room temperature and cryogenic temperatures, and with various pump powers. We find that h-BN quantum emitters exhibit negative Mandel Q in the pulsed excitation regime, as expected. To show the credibility of our methodology, we compared this to Mandel Q measurements for coherent and thermal light sources. Photon statistics for narrow linewidth hBN emitters reveal essentially no dependence of Mandel Q on temperature, within measurement error. Results for pulsed excitation were compared with an extended Jaynes-Cummings model, validating the observed Mandel Q values. Optical pump power can act as a dial to tune the CW Mandel Q by identify pumping conditions yielding the photon statistics with the smallest contributions from noise and experimental measurement errors. We foresee that Mandel Q can serve as a useful parameter to assess the performance of quantum emitters in random number generation, and potentially for other quantum applications.

**Methods:**

**Sample Fabrication:** h-BN crystals purchased from HQGraphene were exfoliated onto 285nm $SiO_2$ on Si substrate (MTI). The sample was annealed in vacuum (1 torr) under argon gas at 850C for 30 minutes.

**RT optical characterization:** We used a homebuilt scanning confocal microscope equipped with an optical spectrometer (Princeton instruments), an HBT setup with a 50:50 beam splitter and two SPADs (MPD) with dead time of around 80 ns, time tagging electronics (Picoharp300), and a wire grid linear polarizer (Thorlabs). We used a CW 532nm laser (coherent) and pulsed 532nm laser (Pico-Quant) for excitation. The setup is explained in more detail in previous publications [22,32].

**Cryo Microscope:** We used an Attodry800 (Attocube) closed loop He cryostat with an optical head. This setup is connected to spectrometer and HBT setup, We use a tunable CW dye laser (Sirah Matisse 2) for PLE. Our PLE methodology is explained in more detail in a previous publication [10].

**Data Analysis:** The data is analyzed by MATLAB and plotted by MATLAB and MS Excel.

**Simulations:** We have developed Boeing's Quantum Design Studio (QDS), a system of procedures and algorithms that aids in definition, solution and analysis of large scale, high-dimensional Hilbert spaces. QDS automates much of the Hamiltonian definition, initial state preparation, ODEs solution, and subsequent expectation value calculations.

**Figures:**

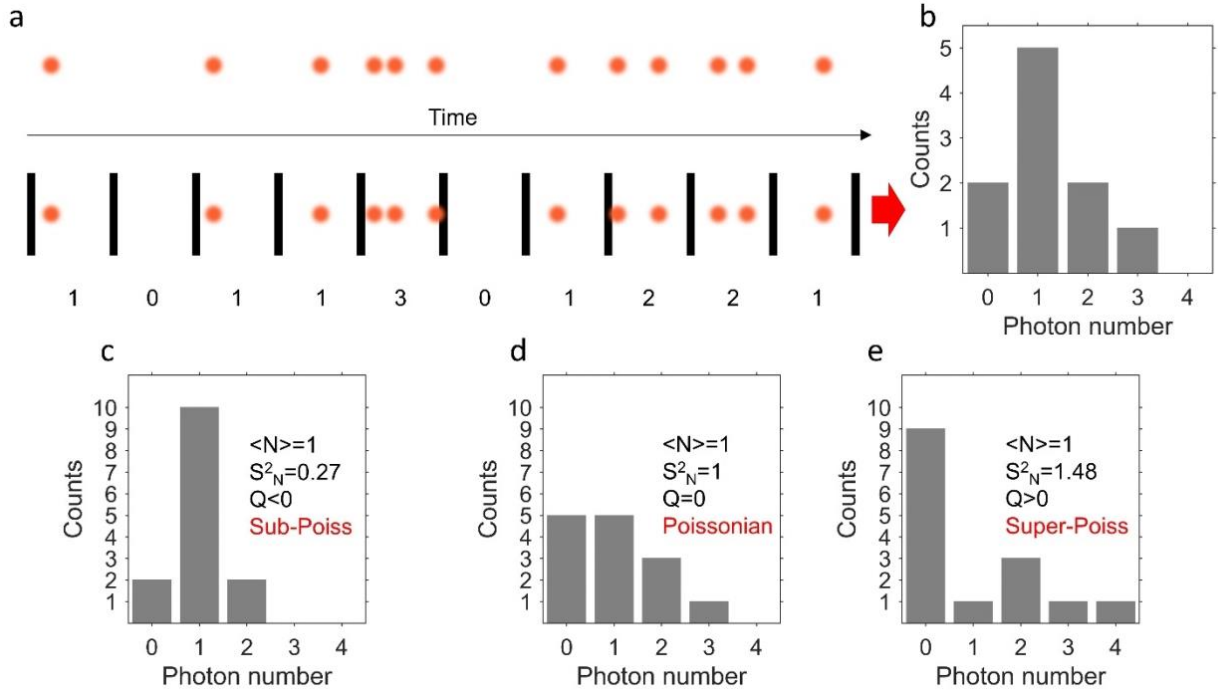

**Figure 1: Schematic of Mandel Q analysis** depicting in (a), a stream of photons with and without division into time bins; in (b), the histogram created by binning the photon stream; in (c-d-e), histograms of sub-Poissonian, Poissonian and super Poissonian photon statistics. The value of Mandel Q depends on whether variance of the distribution is greater or smaller than the average.

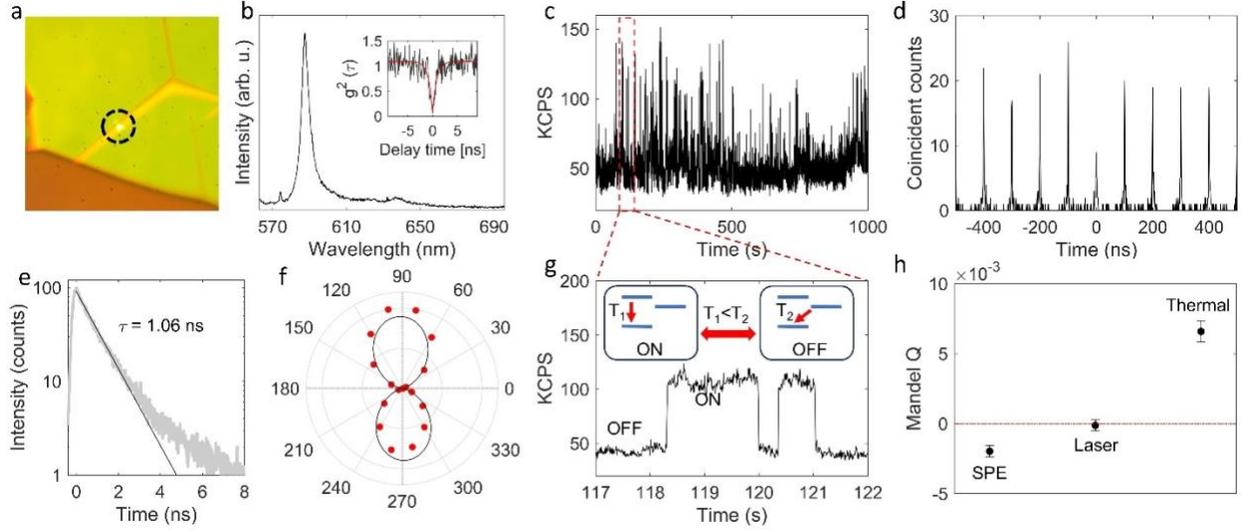

**Figure 2: Characteristics of an h-BN emitter under pulsed excitation**. In (a), microscope image of the h-BN flake containing the quantum emitter; in (b), PL spectra of the emitter with a ZPL at 588nm and a sideband at 638nm. Inset shows CW $g^{(2)}$ function of this emitter with $g^{(2)}(0)=0.07$. In (c & g), emission rate as a function of time. The emitter exhibits two emissive states of on and off with 50kHz and 100kHz emission rate, respectively. In (d), result of pulsed coincidence measurement of the emitter; in (e), TRPL measurement shows a decay lifetime of 1.06 ns; in (f), polarized emission measurement shows a dipole pattern. In (h), single pulse Mandel Q values for emitter in hBN, coherent source and thermal source.

| Single Photon Emitter | | | |
|---|---|---|---|
| # | 1-photon /pulse | 2-photon counts | Mandel Q |
| 1 | 0.005088 | 69 | -0.00238 |
| 2 | 0.005667 | 97 | -0.00224 |
| 3 | 0.005026 | 95 | -0.00125 |
| 4 | 0.005349 | 97 | -0.00172 |
| 5 | 0.004845 | 63 | -0.00224 |
| Average | 5.20E-03 | 84.2 | -1.97E-03 |
| Std. Dev. | 0.000286 | 17 | 4.24E-04 |

| Laser | | | |
|---|---|---|---|
| # | 1-photon /pulse | 2-photon counts | Mandel Q |
| 1 | 0.00494 | 118 | -0.00017 |
| 2 | 0.00497 | 104 | -0.00078 |
| 3 | 0.00497 | 103 | -0.00083 |
| 4 | 0.00491 | 133 | 0.000505 |
| 5 | 0.00492 | 125 | 0.00016 |
| 6 | 0.00493 | 124 | 9.98E-05 |
| 7 | 0.00491 | 123 | 9.67E-05 |
| 8 | 0.00495 | 115 | -0.00031 |
| 9 | 0.00493 | 125 | 0.000133 |
| 10 | 0.00489 | 116 | -0.00015 |
| Average | 4.93E-03 | 118.6 | -1.25E-04 |
| Std. Dev. | 2.60E-04 | 9.5 | 4.01E-04 |

| Thermal source | | | |
|---|---|---|---|
| # | 1-photon /pulse | 2-photon counts | Mandel Q |
| 1 | 5.50E-04 | 18 | 5.99E-03 |
| 2 | 5.46E-04 | 20 | 6.78E-03 |
| 3 | 5.47E-04 | 21 | 7.13E-03 |
| 4 | 5.55E-04 | 17 | 5.58E-03 |
| 5 | 5.49E-04 | 22 | 7.47E-03 |
| Average | 5.49E-04 | 20 | 6.59E-03 |
| Std. Dev. | 3.36E-06 | 2 | 7.68E-04 |

**Table 1:** Comparison of single photon emitter, laser, and thermal light Mandel Q

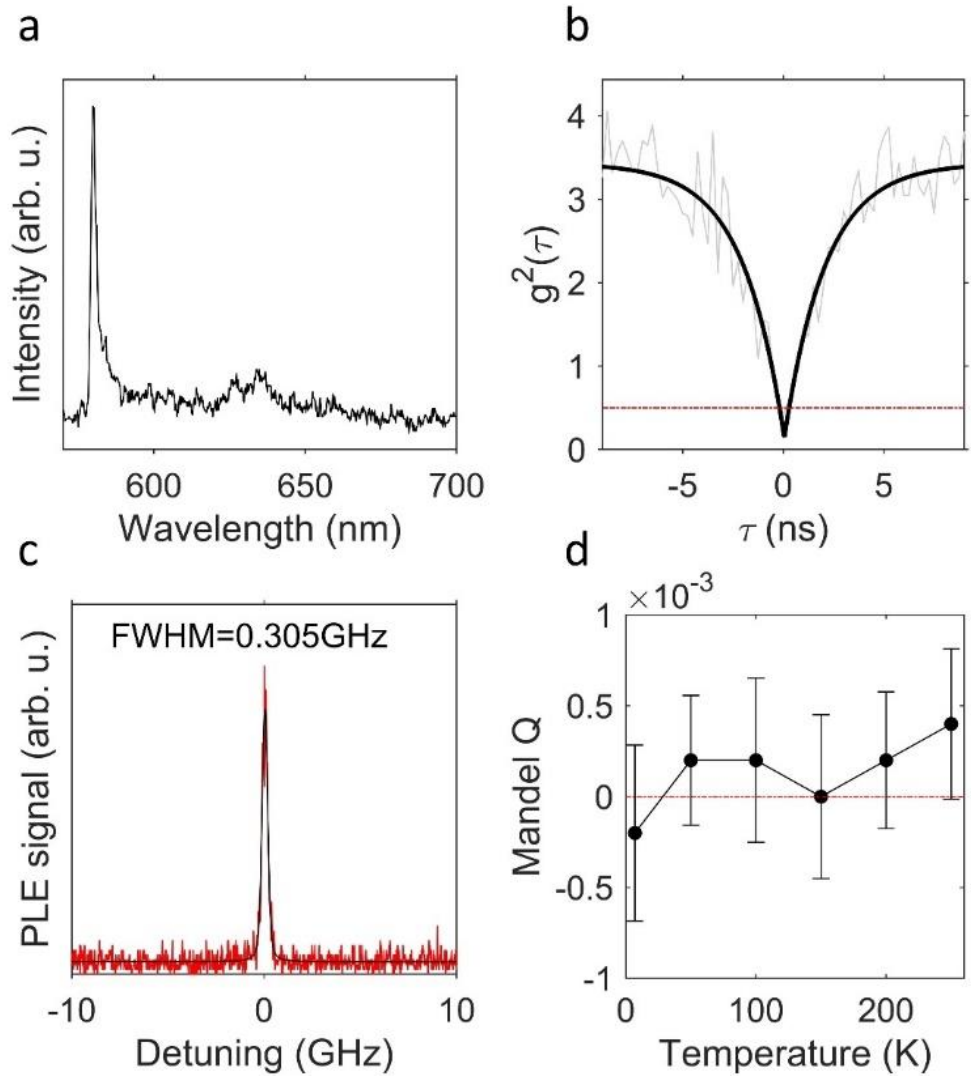

**Figure 3: Temperature dependence of hBN emitter Mandel Q under pulsed excitation**. In (a), PL spectra of the emitter at 7K; in (b), CW $g^{(2)}$ function of the emitter. In (c), result of PLE measurement of the emitter zero phonon line; the emission shows a narrow linewidth of 305MHz. In (d), Mandel Q of emitter at different temperatures from 7K to 250K

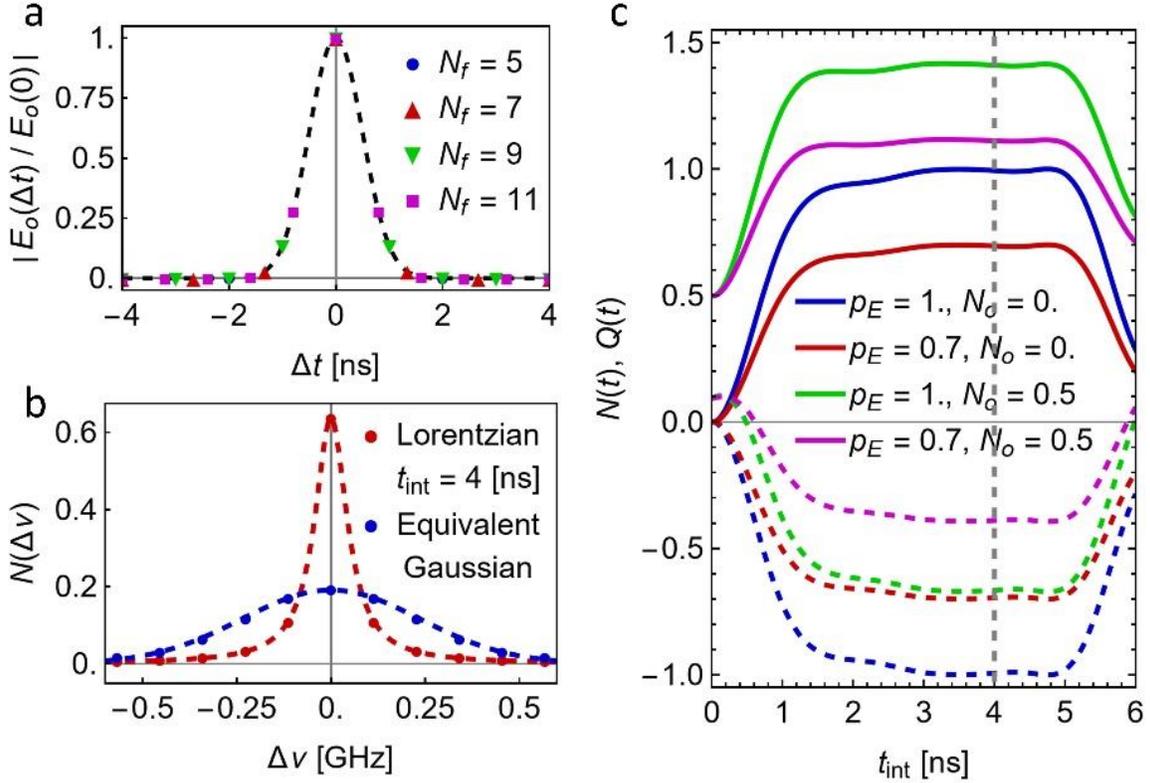

**Figure 4: Results from extended Jaynes-Cummings Model calculations.** (a) Initially presumed Gaussian field emission distribution with $\sigma = 0.5$ ns, periodic duration of 8 ns, from which the modal frequencies are defined; discrete times shown for several number of modes $N_f$. (b) Photon mode occupation per discrete frequency ($N_f = 11$) for (blue) the initially presumed Gaussian distribution and (red) much narrower modal results taken at $t_o = 4$ ns interaction time cutoff, well-fitted as a Lorentzian; notice the isolated pulse interaction stability in the results, particularly between 3 and 5 ns prior to model periodicity effects. (c) Pulse photon statistics vs. interaction time showing (solid) average photon emission $N(t)$ and (dashed) Mandel Q parameter $Q(t)$ for various cases of emission efficiency $p_E$ and initial average thermal noise photons $N_o$ per emission, distributed over $N_f = 5$ modes.

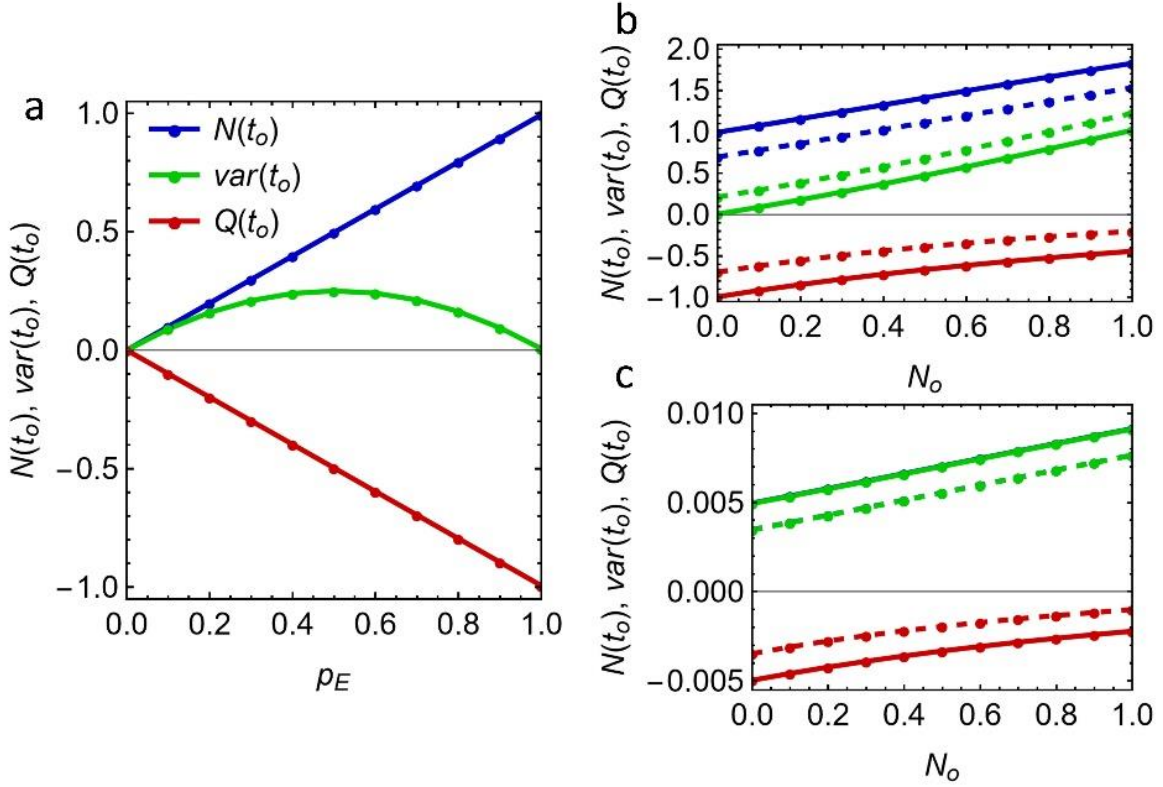

**Figure 5: Results from extended Jaynes-Cummings Model parametric calculations**. In (a), photon statistics for interaction time cutoff $t_o$ (see Fig. 4-c) with no initial noise photons ($N_o = 0$) vs. emission efficiency $p_E$. In (b), photon statistics at $t_o$ vs. $N_o$ for (solid) $p_E = 1$ and (dashed) $p_E = 0.7$. In (c), same results as in (b) but after system loss transformation through (T:R) = (0.005, 0.995) beam splitter; here, $N(t_o)$ is only slightly larger than $var(t_o)$, resulting in small-valued negative $Q$; note the calculated $Q$ values are commensurate with the measurements.

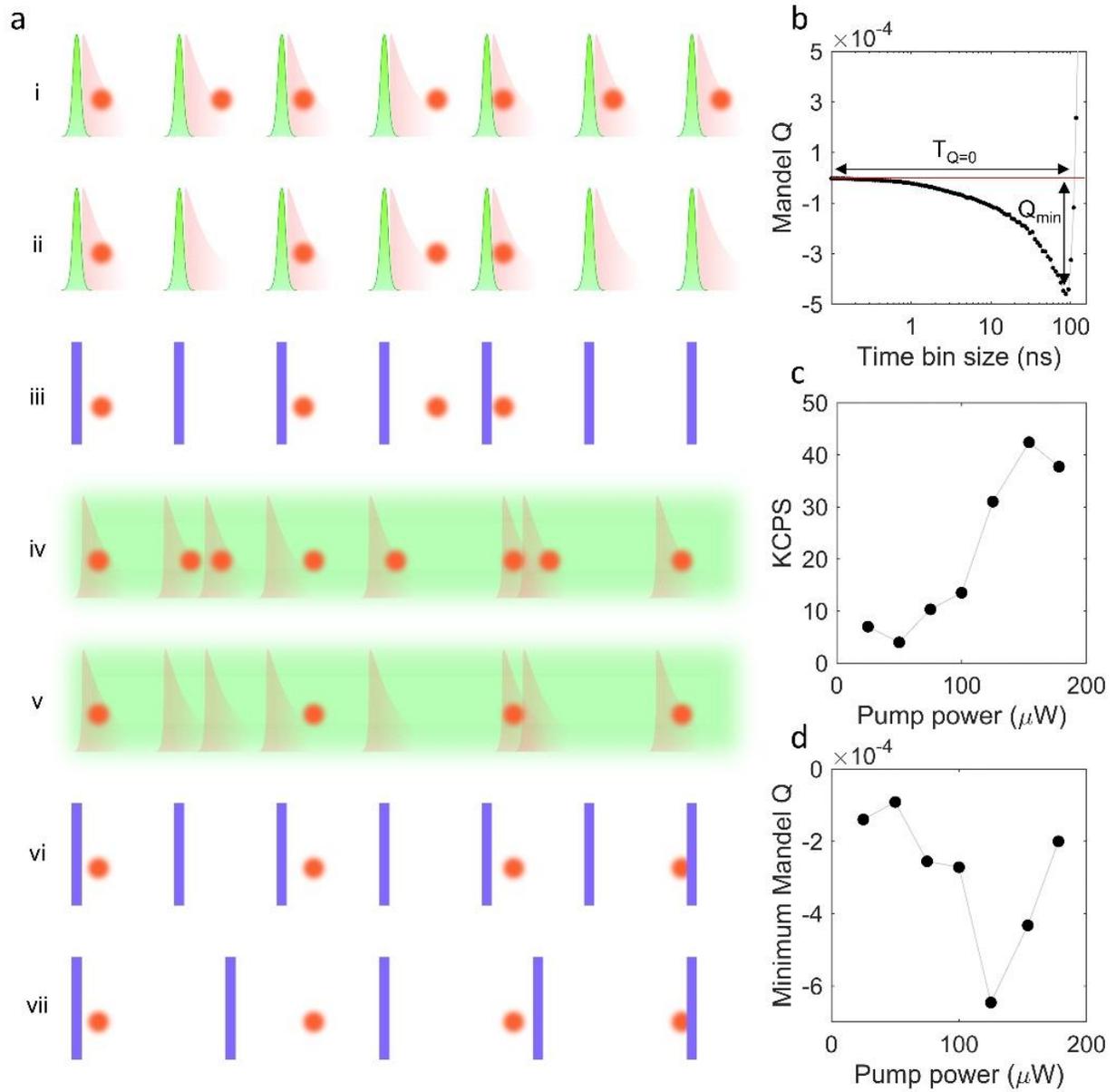

**Figure 6: CW Mandel Q of h-BN emitter.** In (a), schematic of pulsed and CW Mandel Q: (i) ideal case of pulsed excitation will result in one single photon per pulse, (ii) considering losses some of the single photons cannot be detected, (iii) the pulse-to-pulse time scale imposes a natural time scale for binning and reporting the photon statistics; (iv) with CW excitation there is a stream of photons generated based on random excitation and decay of the emitter, (v) due to losses some of the photons generated cannot be detected, (vi-vii) in the case of CW the freedom to choose the bin size can offset the effect of collection efficiency to some extent. In (b), CW Mandel Q as a function of time bin size. In (c), dependence of photon emission rate with pump power indicates operation far from pump power saturation. In (d), CW Mandel Q Vs. pump power. This plot does not show a clear dependence of CW Mandel Q on pump power mainly due to blinking.

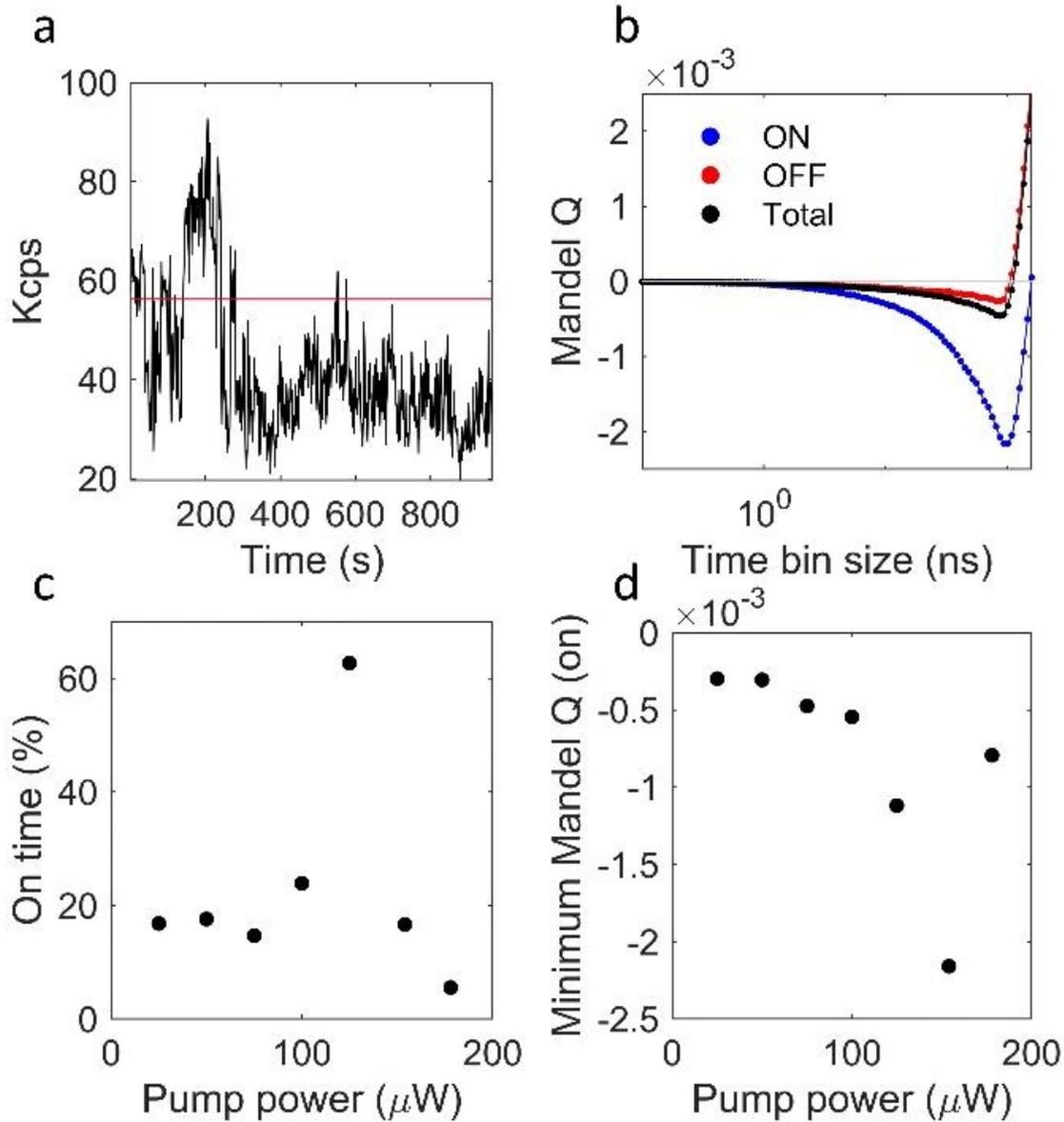

**Figure 7: Effect of blinking on Mandel Q**. (a) Emission rate of the emitter vs. time shows clear on and off states. In (b), dividing data into on and off parts and calculating, Mandel Q for each part. In (c), dependence of blinking on pump power. As the pump power changes, the blinking pattern also changes and there is no clear correlation between pump power and the percentage of on state emission. However, for the maximum power (175μW) the emitter is mostly in the off state. In (d), dependence of minimum CW Mandel Q on pump power. As power increases the Mandel Q decreases.

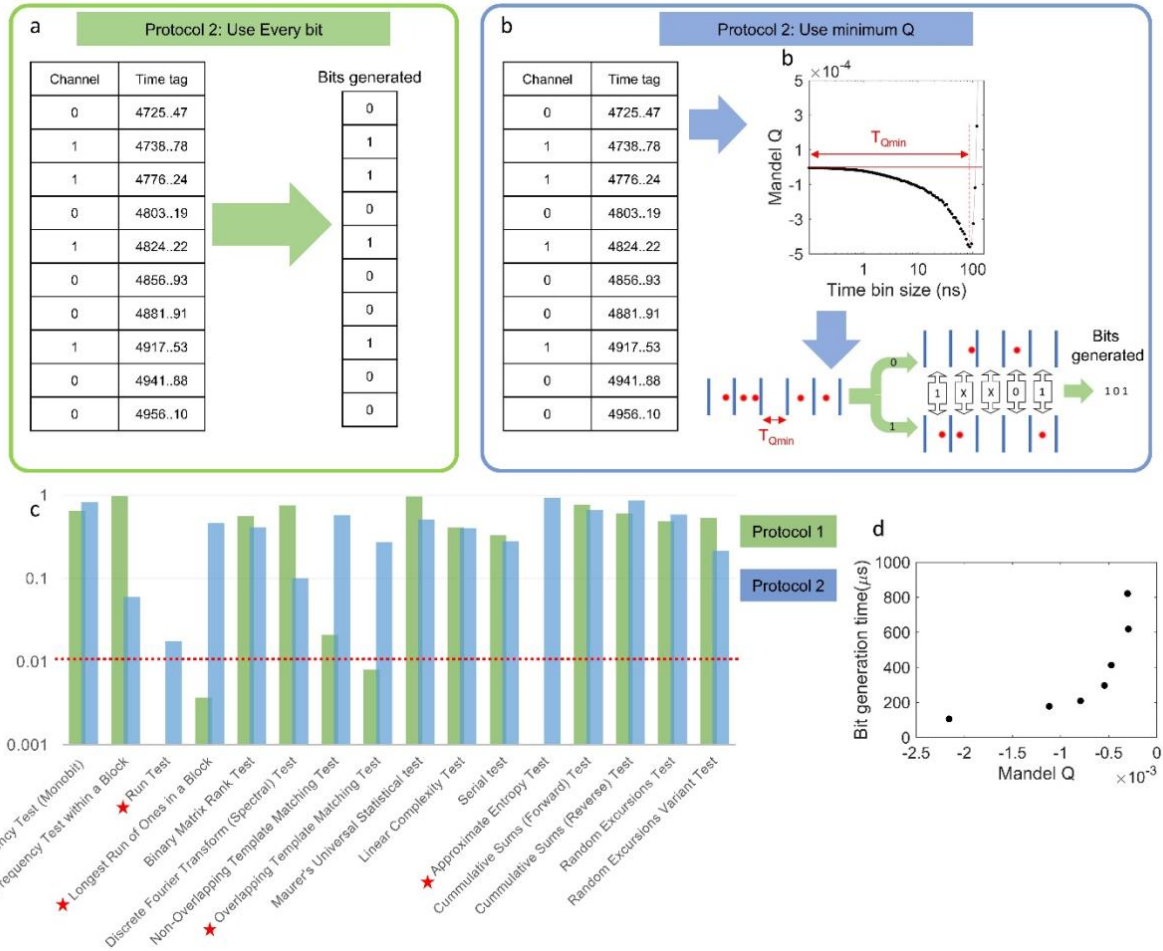

**Figure 8: Random number generation (RNG) with h-BN emitters**. In (a), protocol 1 of RNG. where Mandel Q is not used, but every photon is detected. In (b), protocol 2 of RNG in which the data is binned with a time bin size that ensures the minimum Q value. (c) Comparison of randomness of the bits generated by each protocol via NIST randomness test suite. Protocol 2 passes all tests while protocol 1 does not pass 4 tests. (d) Dependence of time required to generate a bit with protocol 2 with emitter pumped at various pump powers on the Mandel Q. The trend indicates that at more negative Mandel Q values, less time is needed to generate a random bit.

**Supporting information for: Photon statistics analysis of h-BN quantum emitters with pulsed and continuous-wave excitation**

**Authors:** Hamidreza Akbari[1], Pankaj K. Jha[1,2], Kristina Malinowski[1], Harry A. Atwater[1]

[1]*Thomas J. Watson laboratory of Applied Physics, California Institute of Technology, Pasadena, CA., 91125*

[2] *Department of Electrical Engineering and Computer Science, Syracuse University, Syracuse, NY., 13244*

**Details of random number generation:**

We use the HBT setup with a 50:50 beam splitter and picoharp300 records the data in the TTTR mode as a ptu file. we use the code provided online [1] to turn the .ptu file into timetags and channels readable by MATLAB.

Protocol 1: We developed a MATLAB code to go through the data from .ptu file and only look at the channel that recorded the data. The bits are same as the records of channels.

Protocol 2: after performing the CW Mandel Q analysis we use a custum MATLAB code to perform the binning with $T_{Qmin}$ and only choose time bins with one record. And discard other bins. To check randomness we created a set of 20 digit binaries and turned them into decimal the result should be a random decimal between 0 and $2^{20} \sim 10^6$ .we then plotted the histograms of these decimals and the result for all pump powers looked like the uniform distribution.

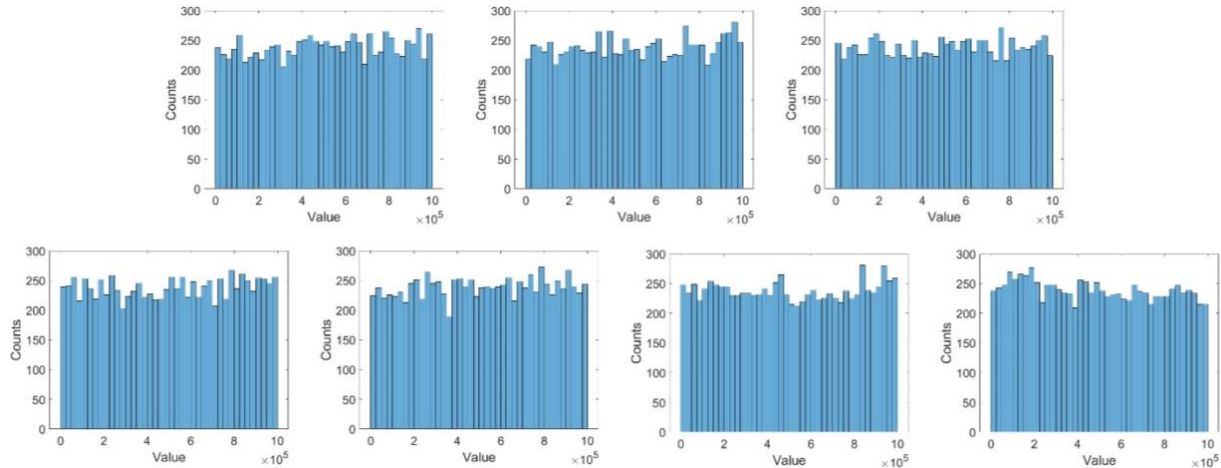

Fig S-1: Histograms of random decimals generated binned from 0 to $10^6$ for powers from 25-175μW

We also performed the NIST randomness test for all of the powers and all of them passed all the 16 tests. The test was performed by the code provided online [2]